\documentclass[prl,twocolumn]{revtex4}
\usepackage{amsfonts}

\usepackage{amsmath}
\usepackage{amsthm}
\usepackage{graphics}
\usepackage{graphicx}
\usepackage{subfigure}
\usepackage{amssymb}
\usepackage{graphics}
\usepackage{graphicx}
\usepackage{color}
\usepackage{subfigure}

\def\k{\textbf{k}}

\begin{document}

\title{Cascades, thermalization and eddy viscosity in helical Galerkin truncated Euler
flows}
\author{G. Krstulovic$^1$, P.D. Mininni$^{2,3}$, M.E. Brachet$^{1,3}$, and A. Pouquet$^3$}
\affiliation{$^1$ Laboratoire de Physique Statistique de l'Ecole Normale Sup\'erieure, associ\'e au CNRS et aux Universit\'es Paris VI et VII, 24 Rue Lhomond, 75231 Paris, France \\
             $^2$ Departamento de F\'\i sica, Facultad de Ciencias Exactas y Naturales, Universidad de Buenos Aires, Ciudad Universitaria, 1428 Buenos Aires, Argentina \\
             $^3$ NCAR, P.O. Box 3000, Boulder, Colorado 80307-3000, U.S.A.}

\begin{abstract}
The dynamics of the truncated Euler equations with helical initial
conditions are studied. Transient energy and helicity cascades
leading to Kraichnan helical absolute equilibrium at small scales
are obtained for the first time. The results of [Cichowlas et al.
Phys. Rev. Lett. 95, 264502 (2005)] are extended to helical flows.
Similarities between the turbulent transient evolution of the ideal
(time-reversible) system and viscous helical flows are found. The
observed differences in the behavior of truncated Euler and
(constant viscosity) Navier-Stokes are qualitatively understood
using the concept of eddy viscosity. The large scales of truncated
Euler equations are then shown to follow quantitatively an effective
Navier-Stokes dynamics based on a variable (scale dependent) eddy
viscosity.
\end{abstract}

\maketitle
\bigskip

The role played by helicity in turbulent flows is not completely
understood. Helicity is relevant in many atmospheric processes, such
as rotating convective (supercell) thunderstorms, the predictability
of which may be enhanced because of its presence \cite{heli}.
However helicity, which is a conserved quantity in the three
dimensional Euler equation, plays no role in the original theory of
turbulence of Kolmogorov. Later studies of absolute equilibrium
ensembles for truncated helical Euler flows by Kraichnan
\cite{KRA73} gave support to an scenario where in helical turbulent
flows both the energy and the helicity cascade towards small scales
\cite{Helcas-BFLLM}, a phenomena recently verified in numerical
simulations \cite{BorueOrszag97,Eyink03,MininniPouquet06}. The
thermalization dynamics of the non-helical spectrally truncated
Euler flows were studied in \cite{CBDB-echel}. However, Kraichnan
helical equilibrium solutions were never directly observed in
simulations. Note that the Galerkin truncated non-helical Euler
dynamics was recently found to emerge as the asymptotic limit of
high order hyperviscous hydrodynamics and that bottlenecks observed
in viscous turbulence may be interpreted as an incomplete
thermalization \cite{frisch-2008}.

In this letter we study truncated helical Euler flows, and consider
the transient turbulent behavior as well as the late time
equilibrium of the system.  Here is a short summary of our main
results. The relaxation toward a Kraichnan helical absolute
equilibrium \cite{KRA73} is observed for the first time. Transient
mixed energy and helicity cascades are found to take place while
more and more modes gather into the Kraichnan time-dependent
statistical equilibrium. It was shown in \cite{CBDB-echel} that, due
to the effect of thermalized small-scales, the spectrally truncated
Euler equation has long-lasting transients behaving similarly to the
dissipative Navier-Stokes equation. These results, obtained for
non-helical flows, are extended to the helical case. The concept of
eddy viscosity, as previously developed in \cite{CBDB-echel} and
\cite{GKMEB2fluid}, is used to qualitatively explain differences
observed between truncated Euler and high-Reynolds number (fixed
viscosity) Navier-Stokes. Finally, the truncated Euler large scale
modes are shown to quantitatively follow an effective Navier-Stokes
dynamics based on a (time and wavenumber dependent) eddy viscosity
that does not depend explicitly on the helicity content in the flow.


Performing spherical Galerkin truncation at wave-number $k_{\rm
max}$ on the incompressible ($ \nabla  \cdot {\bf u}=0$) and
spatially periodic Euler equation $ {\partial_t {\bf u}} + ({\bf u}
\cdot \nabla) {\bf u} =- \nabla p$ yields the following finite
system of ordinary differential equations for the Fourier transform
of the velocity ${\bf \hat u}({\bf k})$ (${\bf k}$ is a 3 D vector
of relative integers satisfying $ |{\bf k}| \leq k_{\rm max}$):
\begin{equation}
{\partial_t { \hat u}_\alpha({\bf k},t)}  =  -\frac{i} {2}
{\mathcal P}_{\alpha \beta \gamma}({\bf k}) \sum_{\bf p} {\hat
u}_\beta({\bf p},t) {\hat u}_\gamma({\bf k-p},t),
\label{eq_discrt}
\end{equation}
where ${\mathcal P}_{\alpha \beta \gamma}=k_\beta P_{\alpha
\gamma}+k_\gamma P_{\alpha \beta}$ with $P_{\alpha
\beta}=\delta_{\alpha \beta}-k_\alpha k_\beta/k^2$.

This time-reversible system exactly conserves the energy
$E=\sum_{k}E(k,t)$ and helicity $H=\sum_{k}H(k,t)$, where the energy
and helicity spectra $E(k,t)$ and $H(k,t)$ are defined by averaging
respectively ${\frac1 2}|{\bf \hat u}({\bf k'},t)|^2 \,$ and ${\bf
\hat u}({\bf k'},t)\cdot{\bf \hat \omega}({\bf -k'},t)$ (${\bf
\omega=\nabla\times {\bf u}}$ is the vorticity) on spherical shells
of width $\Delta k = 1$. It is trivial to show from the definition
of vorticity that $|H(k,t)|\leq 2 k E(k,t)$.


We will use as initial condition ${\bf u}_0$ the sum of two ABC
(Arnold, Beltrami and Childress) flows in the modes $k=3$ and $k=4$,
\begin{equation}
{\bf u}_0(x,y,z)={\bf u}_{\rm ABC}^{(3)}(x,y,z)+{\bf u}_{\rm
ABC}^{(4)}(x,y,z)\label{eq:condini}
\end{equation}
where the basic ABC flow is a maximal helicity stationary solution
of Euler equations in which the vorticity is parallel to the
velocity, explicitly given by
\begin{eqnarray}
{\bf u}_{\rm ABC}^{(k)}(x,y,z) &=& \frac{u_0}{k^2} \left\{ \left[B
\cos(k y) +
    C \sin(k z) \right] \hat{x} + \right. {} \nonumber \\
&& {} + \left[A \sin(k x) + C \cos(k z) \right] \hat{y} +
   {} \nonumber \\
&& {} + \left. \left[A \cos(k x) + B \sin(k y) \right]
   \hat{z} \right\}.
\label{eq:ABC}
\end{eqnarray}
The parameters will be set to $A=0.9$, $B=1$, $C=1.1$ and
$u_0=(A^2+B^2+C^2)^{-1/2}(1/3^4+1/4^4)^{-1/2}$. With this choice
of normalization the initial energy is $E=0.5$ and helicity
$H=3\times4\times(3^3+4^3)/(3^4+4^4)=3.24$.

Numerical solutions of equation (\ref{eq_discrt}) are efficiently
produced using a pseudo-spectral general-periodic code
\cite{PabloCode1} with $512^3$ Fourier modes that is dealiased using
the $2/3$ rule \cite{Got-Ors} by spherical Galerkin truncation at
$\k_{\rm max}=170$. The equations are evolved in time using a second
order Runge-Kutta method, and the code is fully parallelized with
the message passing interface (MPI) library. The numerical method
used is non-dispersive and conserves energy and helicity with high
accuracy.

\begin{figure}[h!]
\begin{center} \includegraphics[height=8.0cm]{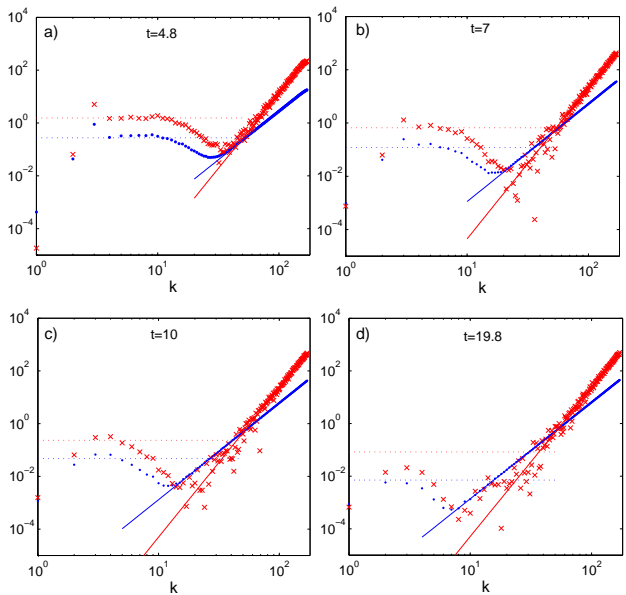}
\caption{Compensated energy ({\tiny $\bullet\bullet\bullet$}) and
helicity spectra ({\tiny $\times\times\times$}) with the predictions
(\ref{HelSpec}) in solid lines and (\ref{eq:speciner}) in dotted
lines. a) $t=4.8$. b) $t=7$. c) $t=10$. d) $t=19.8$.
\label{Fig:speccomp4}} \end{center}
\end{figure}

Fig.\ref{Fig:speccomp4} shows the time-evolution of the energy and
helicity spectra that evolve from (\ref{eq:condini}) compensated by
$k^{5/3}$.
 The plots clearly display a progressive
thermalization similar to that obtained in Cichowlas et
al.\cite{CBDB-echel} but with the non zero helicity cascading to
the right.

The truncated Euler equation dynamics is expected to reach at large
times an absolute equilibrium that is a statistically stationary
gaussian exact solution of the associated Liouville equation
\cite{OrszagAnalytTheo}. When the flow has a non vanishing helicity,
the absolute equilibria of the kinetic energy and helicity predicted
by Kraichnan \cite{KRA73} are
\begin{equation}
  E(k)=\frac{k^2}{\alpha }\frac{ 4 \pi}{1-\beta^2 k^2 /
  \alpha^2}\,;\hspace{0.1cm}
  H(k)= \frac{k^4\beta}{\alpha^2}\frac{ 8 \pi }{1-\beta^2 k^2 / \alpha^2}\label{HelSpec}
\end{equation}
where $\alpha>0$ and $\beta k_{\rm max}<\alpha$ to ensure
integrability. The values of $\alpha$ and $\beta$ are uniquely
determined by the total amount of energy and helicity (verifying
$|H|\leq 2 k_{\rm max} E$) contained in the wavenumber range
$[1,k_{\rm max}]$ \cite{KRA73}.

The final values of $\alpha$ and $\beta$ (when total thermalization
is obtained) corresponding to the initial energy and helicity are
$\alpha=4.12\times 10^7$ and $\beta=7695$. Therefore the
dimensionless number $\beta^2 k^2 / \alpha^2$ is at most of the
order $ 10^{-4}$ and equations (\ref{HelSpec}) thus lead to almost
pure power laws for the energy and helicity spectra, as is manifest
in Fig\ref{Fig:speccomp4}.d. Fig. \ref{Fig:speccomp4} thus shows for
the first time a time evolving helical quasi-equilibrium.


In order to analyze the run in the spirit of Cichowlas et al.
\cite{CBDB-echel} we define $k_{\rm th}(t)$ as the wavenumber where
the thermalized power-law zone starts. We define the thermalized
energy and helicity as
\begin{equation}
{E}_{\rm  th}(t)=\sum_{k_{\rm th}(t)}^{ k_{\rm max}}
E(k,t)\,;\hspace{0.3cm}
 {H}_{\rm th}(t) =  \sum_{k_{\rm th}(t)}^{
k_{\rm max}} H(k,t) \label{Th_energy}
\end{equation}
%
where $E(k,t)$ and $H(t,k)$ are the energy and helicity spectra.

\begin{figure}[h!]
\begin{center}
  \includegraphics[height=7cm]{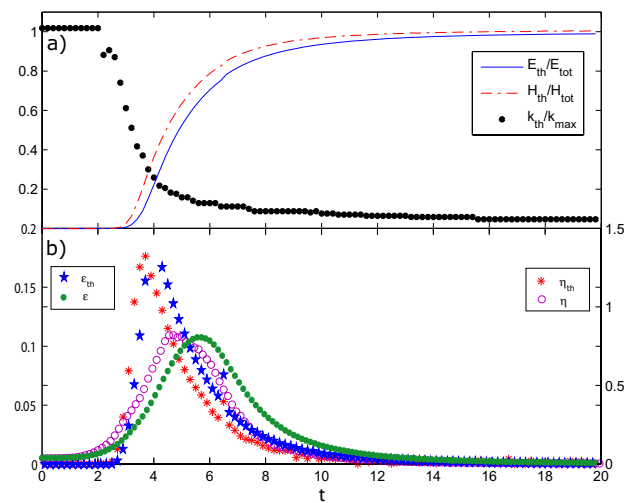}
  \caption{a) Temporal evolution of $E_{\rm th}$ ($-$) ,$H_{\rm th}$ ({\tiny $\cdot-\cdot$}) and $k_{\rm
th}(t)$ ($\cdots$) normalized by their respective initial values.
$E_{\rm tot}=0.5$, $H_{\rm tot}=3.24$ and $k_{\rm max}=170$. b)
Left vertical axis: temporal evolution of $\epsilon_{\rm
th}=\frac{dE_{\rm th}}{dt}$ ({\tiny $\bigstar\bigstar\bigstar$})
and Navier-Stokes energy dissipation
$\epsilon=2\nu_0\sum_{k=1}^{k_{\rm max}} k^2E(k)$ ({\tiny
$\bullet\bullet\bullet$}). Right vertical axis: $\eta_{\rm
th}=\frac{dH_{\rm th}}{dt}$ ($***$) and NS helicity dissipation
$\eta=\nu_0\sum_{k=1}^{k_{\rm max}} k^2H(k)$ ({\small
$\circ\circ\circ$}). }\label{Fig1} \end{center} \end{figure}

 The temporal evolutions of
$E_{\rm th},H_{\rm th}$ and $k_{\rm th}(t)$ are shown in Fig.
\ref{Fig1}.

The values of $\alpha(t)$ and $\beta(t)$ during thermalization can
then be obtained from $E_{\rm th}(t),H_{\rm th}(t)$ and $k_{\rm
th}(t)$ by inverting the system of equations (\ref{Th_energy}) using
$\frac{\beta^2}{\alpha^2} k_{\rm max}^2\ll 1$.

The Kraichnan prediction (\ref{HelSpec}) for the high-$k$ part of
the spectra are shown (in solid lines) in Fig. \ref{Fig:speccomp4}.
The plot shows an excellent agreement with the prediction.

The low-$k$ part of the compensated spectrum presents a flat zone
that amounts to $k^{-5/3}$ scaling for both the energy and helicity
spectra. This $k^{-5/3}$ behavior was predicted by Brissaud et al.
\cite{Helcas-BFLLM} in viscous fluids when there are simultaneous
energy and helicity cascades. The energy and helicity fluxes,
$\epsilon$ and $\eta$ respectively, determine the prefactor in the
inertial range of the spectra:
\begin{equation}
E(k) \sim \epsilon^{2/3}k^{-5/3},\hspace{0.5cm} H(k) \sim
\eta\epsilon^{-1/3}k^{-5/3}\label{eq:speciner} ,
\end{equation}
Helical flows have been also studied in high Reynolds number
numerical simulations of the Navier-Stokes (NS) equation.
Simultaneous energy and helicity cascades leading to the scaling
(\ref{eq:speciner}) have been confirmed when the system is forced at
large scales \cite{BorueOrszag97,Eyink03,MininniPouquet06}.

The energy and helicity fluxes $\epsilon$ and $\eta$ at intermediate
scales in our truncated Euler simulation can be estimated using the
time derivative of the thermalized energy and helicity:
$\epsilon_{\rm th}=\frac{dE_{\rm th}}{dt}$ and $\eta_{\rm
th}=\frac{dH_{\rm th}}{dt}$, whose temporal evolutions are shown in
Fig.  \ref{Fig1}. The predictions (\ref{eq:speciner}) for the
low-$k$ part of the spectra are shown (in dotted lines) in Fig.
\ref{Fig:speccomp4}. The plot shows a good agreement with the data.
Note that Fig. \ref{Fig:speccomp4}.a corresponds to $t=4.8$, that is
just after the time when both the maximum energy and helicity fluxes
(to be interpreted below as ``dissipation'' rates of the
non-thermalized components of the energy and the helicity) are
achieved, see Fig. \ref{Fig1}. In this way $E_{\rm th} $ and $H_{\rm
th}$ determine the thermalized part of the spectra while their time
derivative determines an inertial range.


We now compare the dynamics of the truncated Euler equation with
that of the unforced high-Reynolds number NS equation (i.e.
Eq.(\ref{eq_discrt}) with  $-\nu_0 k^2{ \hat u}_\alpha({\bf k},t)$
added in the r.h.s.) using the initial condition (\ref{eq:condini}).
The viscosity is set to $\nu_0=5\times10^{-4}$, the smallest value
compatible with accurate computations using $k_{\rm max}=170$. A
behavior qualitatively similar to that of the truncated Euler
equation is obtained (see Fig. \ref{Fig1}b). However, the maxima of
the energy and helicity fluxes (or dissipation rates) occur later,
and with smaller values.

We refered above to ``dissipation'' in the context of the ideal
(time-reversible) flow. A proper definition of dissipation in the
truncated Euler flow is now in order. Thermalized modes in truncated
Euler are known to provide an eddy viscosity $\nu_{\rm{eddy}}$ to
the modes with wavenumbers below the transition wavenumber
\cite{CBDB-echel}. It was shown in \cite{GKMEB2fluid} that
Monte-Carlo determinations of $\nu_{\rm eddy}$ are given with good
accuracy by the Eddy Damped Quasi-Normal Markovian (EDQNM) two-point
closure, previously known to reproduce well direct numerical
simulation results \cite{BosBertoglioEDQNM}. For helical flows, the
EDQNM theory provides coupled equations for the energy and helicity
spectra \cite{EDQNM-Andre-Lesieur}, in which using (\ref{HelSpec})
in an analogous way to \cite{GKMEB2fluid} we find a very small
correction of $\nu_{\rm{eddy}}$ that depends on the total amount of
helicity and is of order
$\Delta\nu_{\rm{eddy}}/\nu_{\rm{eddy}}\sim\beta k_{\rm max} /
\alpha\sim 10^{-2}$. Thus the presence of helicity does not affect
significantly the dissipation at large scales and can be safely
neglected in the eddy viscosity expressions. This eddy viscosity has
a strong dependence in $k$ and can also be obtained, in the limit
$k/k_{\rm max}\rightarrow 0$, from the EDQNM eddy viscosity of
Lesieur and Schertzer \cite{LesieruSchertezerEDQNMExpa} using here
an energy spectrum $E(k)\sim k^2$. The result reads
\begin{equation}\label{eq_nuEDQNM}
\nu_{\rm{eddy}}=\frac{\sqrt{E_{\rm{th}}}}{k_{\rm{max}}}\frac{7}{\sqrt{15}\lambda},
\end{equation}
with $\lambda=0.36$. The eddy viscosity $\nu_{\rm{eddy}}$ is thus an
increasing function of time, see $E_{\rm{th}}(t)$ in Fig.
\ref{Fig1}.

\begin{figure}[h!]
\begin{center}
\includegraphics[height=8cm]{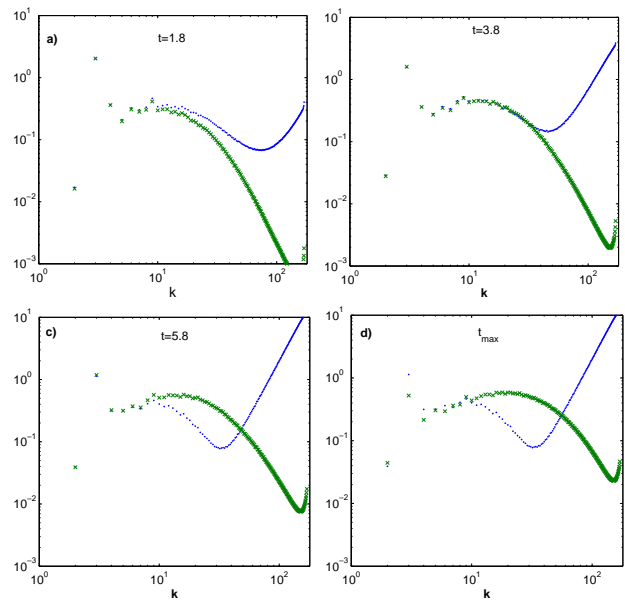}
\caption{Compensated energy spectra of truncated Euler ({\small
$\cdots$}) and Navier-Stokes ({\tiny $\times\times\times$}). a)
$t=1.8$, b) $t=3.8$, c) $t=5.8$. d) Maximum energy-dissipation
time ($t=4.4$ for truncated Euler and $t=5.6$ for NS).
\label{Fig:speccompNS}}
\end{center}
\end{figure}

The time-evolution of truncated Euler and Navier-Stokes spectra
are compared in Fig. \ref{Fig:speccompNS}. At early times the
value of $E_{\rm{th}}$ is very small and therefore the NS
viscosity $\nu_0$ is larger than $\nu_{\rm{eddy}}$, as manifested
by the NS dissipative zone in Fig. \ref{Fig:speccompNS}.a.
As $E_{\rm{th}}(t)$ increases, both viscosities became equal
($t=2.7$). Later, at $t=3.8$, the Navier-Stokes spectrum crosses
the truncated Euler one (Fig. \ref{Fig:speccompNS}b).
The eddy viscosity $\nu_{\rm{eddy}}$ is then much larger than
$\nu_0$ and the truncated Euler dissipative zone lies below the NS
one, see Fig. \ref{Fig:speccompNS}c.
This behavior is also conspicuous when the spectra are compared at
maximum energy-dissipation time ($t=4.4$ for truncated Euler and
$t=5.6$ for NS), see Fig. \ref{Fig:speccompNS}d.


The variation in time of $\nu_{\rm{eddy}}$ thus explains
qualitatively the different behavior of the truncated Euler and
Navier-Stokes spectra in Fig. \ref{Fig:speccompNS}. We now proceed
to check more quantitatively the validity of an effective
dissipation description of thermalization in truncated Euler. To
wit, we introduce an effective Navier-Stokes equation for which the
dissipation is produced by an effective viscosity that depends on
time and wavenumber.

We will use the effective viscosity obtained in \cite{GKMEB2fluid}
which is consistent with both direct Monte-Carlo calculations and
EDQNM closure and is explicitly given by $$
\nu_{\rm{eff}}(k)=\nu_{\rm{eddy}}e^{-3.97k/k_{\rm{max}}},$$ with
$\nu_{\rm{eddy}}$ given in Eq. (\ref{eq_nuEDQNM}).

We thus integrate Eq. (\ref{eq_discrt}) with the viscous term
$-\nu_{\rm{eff}}(k) k^2{ \hat u}_\alpha({\bf k},t)$ added in the
right hand side. The parameter $E_{\rm{th}}$ that fixes the eddy
viscosity in Eq. (\ref{eq_nuEDQNM}) is evolved using the effective
NS dissipation by
\begin{equation}
\frac{d E_{\rm{th}}}{dt}=\sum_{k=1}^{k_{\rm max}}
2\nu_{\rm{eff}}(k)k^2E(k).
\end{equation}
This ensures consistency between the effective NS dissipated
energy and the truncated Euler thermalized energy that drives
$\nu_{\rm{eddy}}$.

To initialize the effective NS equation we integrate the truncated
Euler equation (\ref{eq_discrt}) with the initial condition
(\ref{eq:condini}) until the $k^2$-thermalized zone is clearly
present ($t=4.77$). The value of $E_{\rm{th}}$ is then computed
using equations (\ref{Th_energy}). The low-passed velocity ${\bf
u}^<$, defined by $${\bf u}^<(\textbf{r}) = \sum
\frac{1}{2}\left(1+\tanh{\left[2({|\k|-k_{\rm th}})\right]}\right)
\hat{{\bf u}}_\k e^{i\k\cdot\textbf{r}}$$ is used as initial data
for the effective Navier-Stokes dynamics.

\begin{figure}[h!]
\begin{center}
  \includegraphics[height=9.2cm]{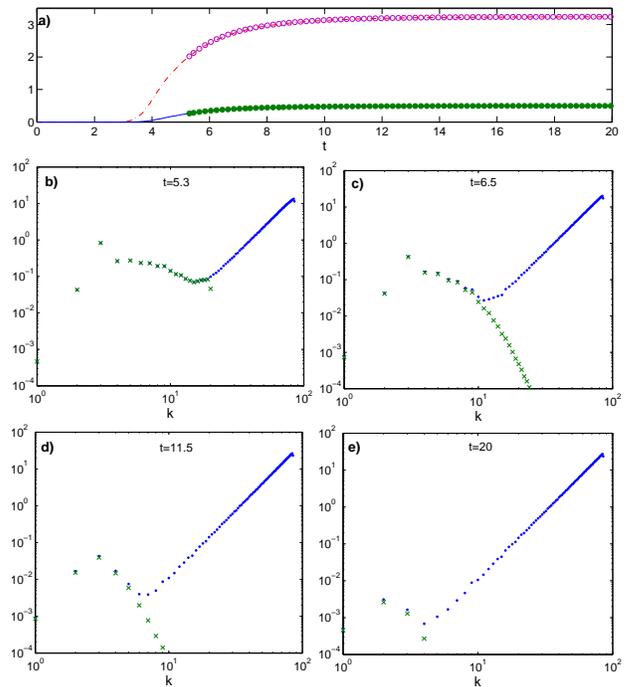}
  \caption{Effective NS run with $k_{\rm max}=85$. a) Temporal evolution of $E_{\rm th}$ ($-$), $H_{\rm th}$ ({\tiny $\cdot-\cdot$}) from truncated
  Euler, energy ({\tiny $\bullet\bullet\bullet$}) and helicity ({\small $\circ\circ\circ$}) from effective NS.
     b-e) Temporal evolution of compensated energy spectra of truncated Euler ({\small
$\cdots$}) and effective Navier-Stokes ({\tiny
$\times\times\times$}) .}\label{Fig5}
\end{center}
\end{figure}

Results of a truncated Euler and effective NS with $k_{\rm
max}=85$ are shown in Fig. \ref{Fig5}. In Fig. \ref{Fig5}.a the
energy and helicity dissipated in effective NS [$E_{\rm tot}-E(t)$
and $H_{\rm tot}-H(t)$ respectively] are compared to $E_{\rm th}$
and $H_{\rm th}$  showing a good agreement. Next, the temporal
evolution of both energy spectra from the initial time $t=5.3$
(Fig. \ref{Fig5}.b) to $t=20$ (Fig. \ref{Fig5}.e)  is confronted,
demonstrating that the low-$k$ dynamics of truncated Euler is well
reproduce by the effective Navier-Stokes equations.

In summary, we observed the relaxation of the truncated Euler
dynamics toward a Kraichnan helical absolute equilibrium. Transient
mixed energy and helicity cascades were found to take place. Eddy
viscosity was found to qualitatively explain the different behaviors
of truncated Euler and (constant viscosity) Navier-Stokes. The large
scale of Galerkin truncated Euler were shown to quantitatively
follow an effective Navier-Stokes dynamics based on a variable
helicity-independent eddy viscosity. In conclusion, with its
built-in eddy viscosity, the Galerkin truncated Euler equations
appears as a minimal model of turbulence.

\textbf{Acknowledgments:} We acknowledge discussions with U. Frisch.
and J.Z. Zhu. P.D.M. is a member of the Carrera del Investigador
Cient\'{\i}fico of CONICET. The computations were carried out at
NCAR and IDRIS (CNRS).

\bibliographystyle{unsrt}

\end{document}